\documentclass[singlecolumn,showpacs,amsmath,amssymb,amsfonts,nofootinbib,plb]{revtex4}
\usepackage{graphicx}
\usepackage{dcolumn}
\usepackage{bm}
\usepackage{ulem}
\usepackage{overpic}
\usepackage{amssymb}

\newcommand{\nn}{\nonumber}

\def\beq{\begin{equation}}
\def\eeq{\end{equation}}

\usepackage{soul}
\usepackage{cancel}

\usepackage[usenames]{color}

\begin{document}
\input{epsf}

\title{Orbital Angular Momentum Distribution of Gluons at Small-$x$ \\
Analytic interpolation between Ji and Jaffe-Manohar}

\author{Khatiza Banu}

\author{Nahid Vasim}

\author {Raktim Abir}   
\email{raktim.ph@amu.ac.in}

\affiliation{Department of Physics, Aligarh Muslim University, Aligarh - $202001$, India.}

 \begin{abstract}
In this work we first introduced a generalized Wilson line gauge link that reproduces both staple and near straight links along light cone in different parameter limits. We then studied  the gauge invariant bi-local orbital angular momentum operator with such a general gauge link, in the framework of Chen et. al. decomposition of gauge fields. At the appropriate combination of limits, the operator reproduces both Jaffe-Manohar  and Ji's operator structure and offers a continuous analytical interpolation between the two. 
 \end{abstract}


\date{\today}
\maketitle

\section{Introduction}
It has been more than three decades since the discovery of proton {\it spin crisis} by the European Muon Collaboration (EMC) \cite{Ashman:1987hv}. By now, a lot of progress has been achieved both on the theoretical and experimental fronts to understand and validate the proton spin sum rule. Moreover, a resurgence of interests have been triggered in the wake of the planned Electron-Ion Collider (EIC) at Brookhaven National Laboratory (BNL) \cite{Aschenauer:2017jsk}. 
The sum rule for  proton spin contains both helicity contribution and orbital angular momentum contribution each separately for quarks and gluons, as, 
\begin{eqnarray}
S_{q+\bar{q}} + L_{q+\bar{q}} + S_{G} + L_G = \frac{1}{2}~,
\end{eqnarray}
This can often be presented in terms of the corresponding distributions as, 
\begin{eqnarray}
\int_{0}^{1} dx \left[  \Delta \Sigma \left(x,Q^2\right) + \Delta G \left(x, Q^2\right) + L_{q+\bar{q}}\left(x, Q^2 \right) + L_G(x,Q^2)     \right] = \frac{1}{2} ~,
\end{eqnarray}
where $\Delta  \Sigma \left(x,Q^2\right)$, $\Delta G \left(x, Q^2\right)$ are helicity distributions for quarks and gluons whereas $L_{q+\bar{q}}\left(x, Q^2 \right)$ and  $L_G(x,Q^2)$ are that for angular momentum respectively. 
\vspace{0.4cm}

\begin{figure}
\includegraphics[width=9cm,height=6cm]{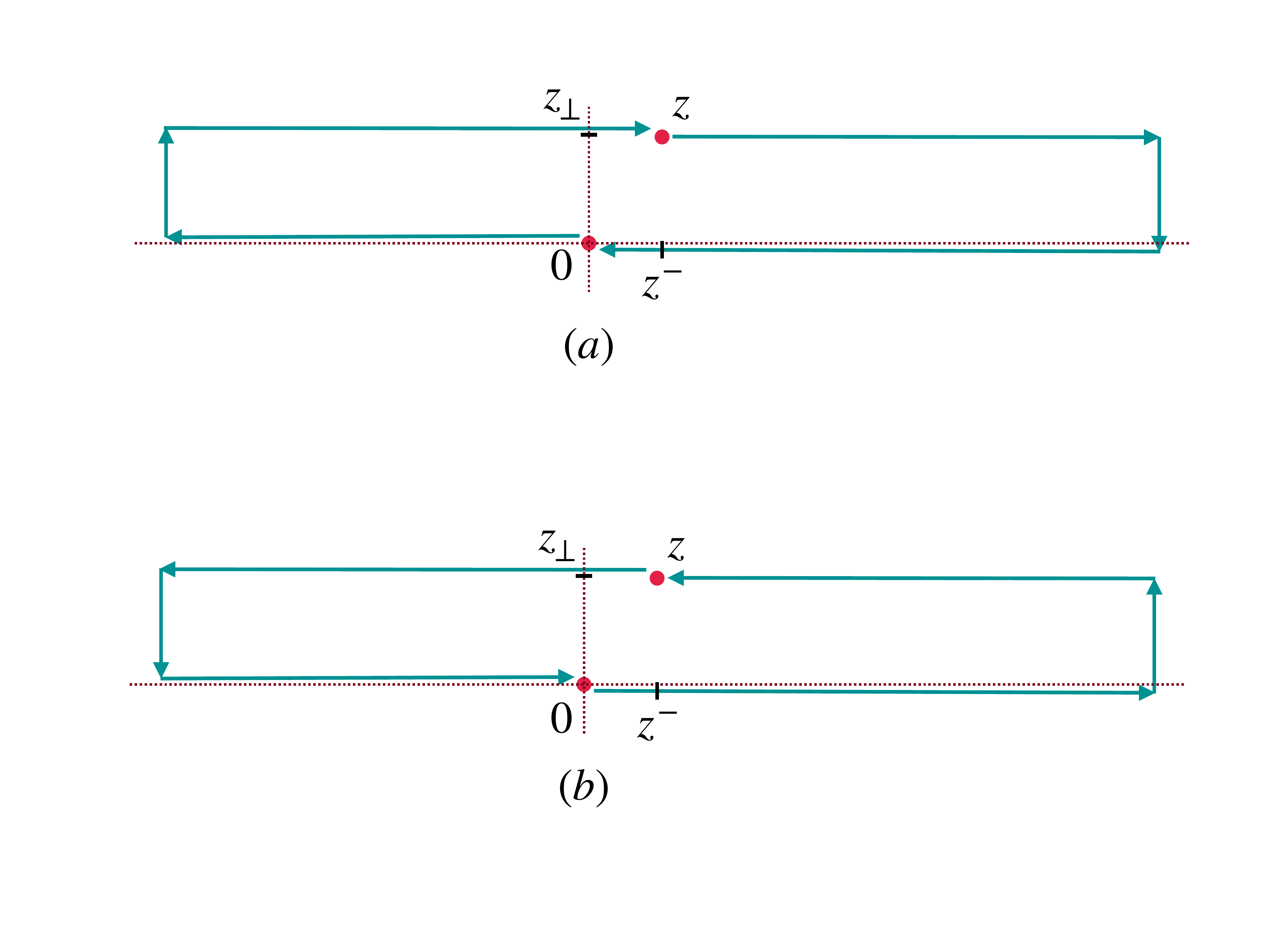}
\caption{The gauge link structure $[\eta_1,\eta_2]$ equals $[-\infty, +\infty^{\dagger}]$ and $[+\infty, -\infty^{\dagger}]$. Both correspond to the dipole type JM orbital angular momentum.}
\label{myfig1}
\end{figure}

\begin{figure}
\includegraphics[width=9cm,height=6cm]{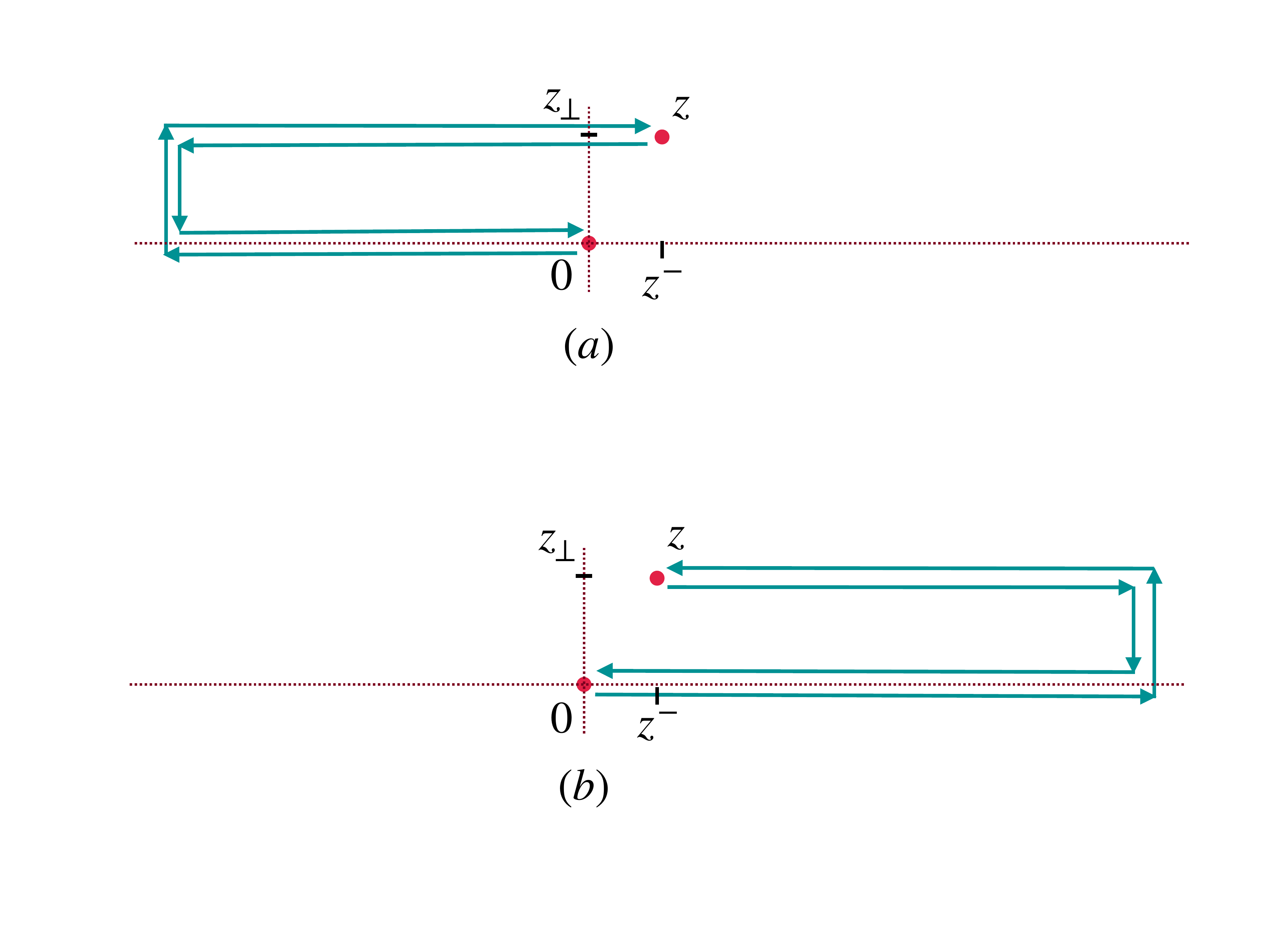}
\caption{The gauge link structure $[\eta_1,\eta_2]$ equals $[-\infty, -\infty^{\dagger}]$ and $[+\infty, +\infty^{\dagger}]$. Both correspond to the  Weizs\"{a}cker-Williams type JM orbital angular momentum.}
\label{myfig2}
\end{figure}

\begin{figure}
\includegraphics[width=9cm,height=6cm]{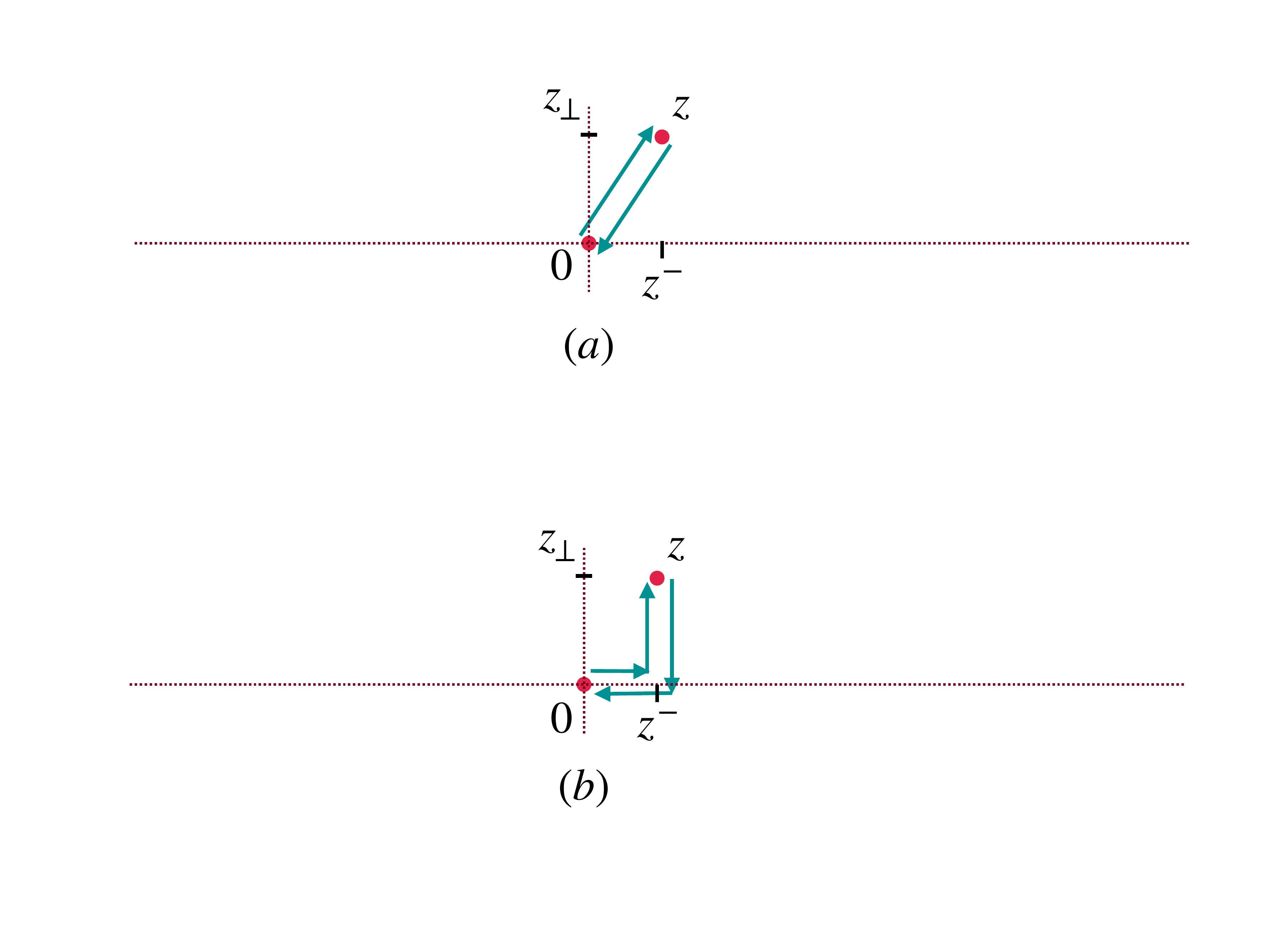}
\caption{The straight gauge link structure $[\eta_1,\eta_2]$ equals $[0,0^{\dagger}]$. This appears in Ji's orbital angular momentum.}
\label{myfig3}
\end{figure}

Gluon orbital angular momentum (OAM) is essentially a two-point correlation function of the field strength tensors at two space-time points. To make such bi-local operators gauge invariant, one needs to introduce appropriate gauge links between the field strength tensors. Two important choices in this regard are, 
 (a) the {\it staple gauge links} and (b) the {\it straight gauge links}. The staple gauge links may show up either as the past pointing or as the future pointing links. This leads to two main types of OAMs namely the dipole type and the Weizs\"{a}cker-Williams type. The dipole OAM contains both the past and future links (Fig[\ref{myfig1}]) while the Weizs\"{a}cker-Williams type OAM contains either the past pointing or the future pointing links along light cone (Fig[\ref{myfig2}]) - nonetheless both of them lead to Jaffe-Manohar \cite{Jaffe:1989jz} definition of OAM operator.
The gauge links connecting the two space-time points could also be the straight one as shown in Fig[\ref{myfig3}], leading to Ji's \cite{Ji:2012sj} definition of OAM.

\vspace{0.4cm}

In this study, we have derived a generalized form of the gauge-invariant OAM operator, presented in Eq.\eqref{Godfather}, which accommodates all possible geometrics of the gauge links. 
%
%
%
%
%
%
%
%
%
In the appropriate limit the operator reproduces both Jaffe-Manohar  and Ji's operator structure and offers a continuous analytical interpolation between the two. Next we jotted down some of the key points that motivate to have  a generalised operator and the continuous interpolation like the one we studied in this work.

\vspace{0.4cm}

{\it Potential OAM - the integrated torque accumulated by a gluon~:} \\

A straight Wilson line gauge link yields Ji orbital angular momentum 
while an infinite staple shaped gauge link yields Jaffe-Manohar 
orbital angular momentum. Unlike the Ji OAM, the Jaffe-Manohar 
decomposition of OAM affected by the final or/and initial state 
interactions. Hence the difference between the two is often interpreted 
as the integrated torque accumulated by a gluon struck in a deep 
inelastic scattering process as it enters or exits the proton, 
through initial or final state interactions - exerted on the 
outgoing quark or gluon by the chromo-magnetic field produced by 
the spectators. The difference between the two OAMs are directly connected 
to the off-forward extension of a Qiu-Sterman term and often cited 
as potential OAM \cite{Burkardt:2012sd}.

\vspace{0.4cm}

{\it Quasi PDF in lattice - connection to TMDs in DIS and Drell-Yan processes~:} 

\vspace{0.4cm}

The operator defining the quasi PDF involves a straight line link from $0$ to $z$ rather than a staple link often appears in the definitions of TMDs to describe the Drell-Yan and semi-inclusive DIS processes. While the staple link reflects initial and final state interactions inherent in these processes, the straight link describes the internal structure of the hadron when it is in the non-disturbed or primordial state. It is unlike that such a TMD can be measured in a scattering experiment, however being a well-defined object they can be measured in the lattice.   As the quasi PDF is defined with space separation connected by straight gauge link a proper synergy of 2d Fourier transform and interpolation like the one addressed in this work can connect it to the TMDs that show up in the scattering processes \cite{Radyushkin:2017cyf}.

\vspace{0.4cm}

{\it TMD PDFs - Weizs\"{a}cker-Williams to dipole~:} 

\vspace{0.4cm}

Unintegrated gluon distribution or transverse momentum-dependent gluon distribution functions are one of the key topics to be fully investigated in the upcoming EIC. The TMD PDFs can either be probed in quark-antiquark jet correlation in DIS (the Weizs\"{a}cker-Williams distribution) or the direct photon-jet correlation in $pA$ collision. The various TMD PDFs involved in different processes specifically in different di jet channels in $pA$ or $eA$ collisions are related by the two universal ones: the Weizs\"{a}cker-Williams TMD and the dipole TMD specifically in small-$x$ at large $N_c$ limit. Both the WW distribution or dipole distribution are essentially dimension four two-point correlation functions of classical gluon fields. The operator definitions are different only by the way the gauge links are oriented. An interpolation of the link, the kind we studied here, lead to a universal structure that can reproduce WW and Dipole TMD at the appropriate limit.

\vspace{0.4cm}

In Section II, we briefly review the essentials of canonical gauge invariant decomposition, of parton angular momentum by Chen et. al., where one has to separate the gauge potential into so called {\it pure} part and {\it physical} or dynamical part. 
In Section III, we studied the transverse derivatives of the staple gauge links along the light cone. Here we introduced two parameters, $\eta_1$ and $\eta_2$ to make the extent of the gauge link arbitrary with the anticipation that at different combinations of the limit of the parameters, it would produce the past pointing, the future pointing, or near straight gauge links. We finally express the transverse derivative of such general gauge link in terms of the pure and physical gauge components. This also explicitly contains the gauge link extent parameters.
In Section IV, we recalled the connection between the orbital angular momentum of gluons and the gluon Wigner distribution and then derive the OAM operators in terms of pure gauge and the extent parameters. 
In Section V, we moved to small-$x$ eikonal limit  and went on to integrate both $z^-$ and $z_\perp$, the position components, which lead to the general gauge invariant OAM operator structure, valid for arbitrary geometry of the gauge link at small-$x$.  
In Section VI, we briefly noted the results for sub-eikonal cases. 
We conclude and give our outlook at the end.

\section{Gauge invariant decomposition of angular momentum}
Decomposition of angular momentum in as gauge invariant way can be achieved by separating the gauge field into the ${\it pure}$  and ${\it physical}$ parts of the field, first proposed by Chen et. al. \cite{Chen:2008ag,Chen:2009mr}, 
\begin{eqnarray}
A_\mu = A_\mu^{\rm pure} + A_\mu^{\rm phys}~.  \label{LaLaLand5}
\end{eqnarray}
Pure gauge field as differential $1$-form defined by identically vanishing the corresponding $2$-form $i.e.$ field strength tensor,  
\begin{eqnarray}
F_{\mu\nu}^{~\rm pure} ~=~  \partial_{\mu}~A_{\nu}^{~\rm pure}  - 
\partial_{\nu}~A_{\mu}^{~\rm pure} - ig \left[A_{\mu}^{\rm pure}, ~A_{\nu}^{\rm pure}\right] ~=~ 0~.
\end{eqnarray}
One can define covariant derivative as well, 
\begin{eqnarray}
D_{\mu}^{\rm~pure} = \partial_\mu - ig A_\mu^{\rm pure}  ~.
\end{eqnarray}
The covariant derivative satisfy the commutation relation that vanishes because field tensor vanishes identically, 
\begin{eqnarray}
\left[D_\mu^{\rm pure}, ~ D_\nu^{\rm pure} \right] & =& - i g F_{\mu\nu}^{\rm pure} ~ = ~ 0 ~.
\end{eqnarray}
Under gauge transformation the pure part of the gauge field transform as, 
\begin{eqnarray}
A_{\mu}^{~\rm pure} \longmapsto A_{\mu}^{~\rm pure}  = U(x)~A_{\mu}^{~\rm pure}~U^{-1}(x)  + \frac{i}{g}~U(x)~\partial_\mu~ U^{-1}(x) ~,
\end{eqnarray}
On the other hand physical part of the gauge field transform as, 
\begin{eqnarray}
A_{\mu}^{\rm phys} ~ \mapsto 
~ A_{\mu}^{\rm phys}  =  U(x) ~ A_{\mu}^{\rm phys} ~ U^{-1}(x)~,
\end{eqnarray}
and makes the field strength tensor non-zero, 
\begin{eqnarray}
F_{\mu\nu} \equiv {\cal D}_{\mu}^{\rm pure} ~  A_{\nu}^{\rm phys}   -   {\cal D}_{\nu}^{\rm pure}  A_{\mu}^{\rm phys}   - ig ~[A_{\mu}^{\rm phys}, ~A_{\nu}^{\rm phys}]~.
\end{eqnarray}
This makes physical part of the gauge field to be dynamical degrees of freedom of the theory. The covariant derivative in the adjoint representation is defined as, 
\begin{eqnarray}
{\cal D}_{\mu}^{\rm ~ pure} ~A_{\nu}^{\rm ~phys} &=& \partial_{\mu}  A_{\nu}^{\rm ~phys}
~ -~ ig \left[A_{\mu}^{\rm~  pure}, ~ A_{\nu}^{\rm ~ phys}\right]  ~.
\end{eqnarray}
One often define the {\it natural gauge} as the gauge where $A_\mu^{~\rm pure} = 0$ and $A_\mu = A_\mu^{~\rm phys}$.
Now this separation of gauge field in terms of pure and physical part is not unique as still some gauge freedom remain. This requires further constrain on $A_{\mu}^{\rm ~ phys}$ which essentially makes it, instead of local, a non-local functional of full gauge field, $e.g.$ in the $A^{+}_{\rm phys}=0 $ gauge, as, 
\begin{eqnarray}
A^{\mu}_{\rm phys, \pm}\left(x^-,~x_\perp\right) ~=~ \int_{\pm \infty^-}^{x^-} d\omega^- ~ U\left(x^-,~\omega^-;~x_\perp \right)~F^{+\mu}\left(\omega^-,~x_\perp\right)
U\left(\omega^-,~x^-;~x_\perp \right),  \label{LaLaLand4}
\end{eqnarray}
where, 
\begin{eqnarray}
U(a^-,~b^-;~x_\perp) ~ \equiv ~ {\cal P}\left(ig\int^{a^-}_{b^-}~dx^-~A^{+}\left(x^-,~x_\perp\right)\right).
\end{eqnarray}
For recent review and a comprehensive study within scalar di-quark model, see \cite{Amor-Quiroz:2020qmw}.

\section{Transverse derivative of staple gauge links with finite extent}
\noindent In this section we review the derivative of a staple gauge link with respect to  transverse coordinate, along light cone, 
\begin{eqnarray}
{\cal U}_{[\eta]}\left(0,~z\right) = 
U\left(0,~\eta^-;~0_\perp\right)
U\left(\eta^-;~   0_\perp, ~z_\perp \right)
U\left(\eta^-,~ z^-; ~z_\perp\right)     
\label{gifted21}
\end{eqnarray}
In the limit $\eta = +\infty^-$ and $\eta = -\infty^-$ the link above becomes past pointing and future pointing staple infinite gauge link along light cone,
\begin{eqnarray}
{\cal U}_{[\eta=-\infty]} &=& U\left(0,-\infty^-;~0_\perp\right)
U\left(-\infty^-;~   0_\perp, ~z_\perp \right)
U\left(-\infty^-,~ z^-; ~z_\perp\right)~,   \\ 
\nn \\
{\cal U}_{[\eta=+\infty]} &=& U\left(0,\infty^-;~0_\perp\right)
U\left(\infty^-;~   0_\perp, ~z_\perp \right)
U\left(\infty^-,~ z^-; ~z_\perp\right) ~. 
\end{eqnarray}
Another interesting limit is when $\eta \rightarrow 0$, which is not exactly the straight gauge link but the staple link with shortest extent. 
This is partly motivated by the recent work by M. Engelhardt et. al.
where the authors considered a general form of staple shaped path and makes an interpolation  between both Ji and Jaffe-Manohar orbital angular momentum when calculating the OAM of quarks on the  lattice using direct derivative method \cite{Engelhardt:2020qtg}.

 \vspace{0.5cm}
\noindent We now recall the Wilson line from $y$ to $z$ along some arbitrary path ${\cal C}$ that can be parametrize via $s^{\mu}$, 
\begin{eqnarray}
U(z,y) &=& ~ P \exp \left(ig \int_{y,~{\cal C}}^{z} ~A_{\mu}\left(s\right)ds^{\mu} \right)~. 
\end{eqnarray}
The position derivative of the Wilson line at any point on the path $\cal C$ is given by, 
\begin{eqnarray}
\left.\frac{\partial}{\partial x^\lambda} U(z,y) ~ = ~ ig~ U(z,s) A_{\mu}(s) ~\frac{\partial s^\mu}{\partial x^\lambda}~U(s,y)\right|_{s=y}^{s=z} ~+ ~ ig\int_{y}^{z} ds^\nu ~U(z,s)~F_{\mu\nu}(s)~\frac{\partial s^\mu}{\partial x^\lambda} ~U(s,y)~.  \label{gifted45}
\end{eqnarray}
Eq.\eqref{gifted45} can be used to derive the derivative of ${\cal U}_{[\eta]}\left(0,~z\right)$ as given in Eq.\eqref{gifted21} with respect to $z_\perp$ as, 
\begin{eqnarray}
&&\frac{\partial}{\partial z_\perp^i} ~ {\cal U}_{~[\eta]}\left(0,~z\right)   \nn \\
&& ~  \nn \\
&=& 
+~ ig ~ U\left(0^-,~\eta^-;~0_\perp\right)
A^{i}\left(\eta^-,~0_\perp \right)
U\left(\eta^-;~0_\perp, ~  z_\perp\right)
U\left(\eta^-,~ z^-;~  z_\perp\right)  \nn \\
&& ~  \nn \\
&& -~ig ~ U\left(0^-,~\eta^-;~0_\perp\right)
U\left(\eta^-;~0_\perp, ~  z_\perp\right)
U\left(\eta^-,~z^-; ~  z_\perp\right)  
A^{i}\left(z^-,~ z_\perp \right)    \nn \\
&& ~ \nn \\
&& +~ i g~ U\left(0^-,~\eta^-;~0_\perp\right) ~ 
\int^{0_\perp}_{z_\perp} d\omega_\perp ~
U\left(\eta^-;~0_\perp,~\omega_\perp\right) ~  F^{\perp i}\left(\eta^-,~\omega_\perp\right) ~ U\left(\eta^-;~\omega_\perp ,~ z_\perp \right) U\left(\eta^-,~z^-;~ z_\perp\right)  \nn  \\
&& ~ \nn \\
&&  +~ ig ~ U\left(0^-,~\eta^-;~0_\perp\right)
U\left(\eta^-;~0_\perp, ~  z_\perp\right)  \int^{ \eta^-}_{z^-}d\omega^- ~ U\left( \eta^-,~\omega^-;~  z_\perp \right)
~ F^{+i}\left(\omega^-,~  z_\perp\right)
~U\left(\omega^-, ~  z^-;~ z_\perp \right)   ~.  \\
&& ~ \nn 
\end{eqnarray}
In the limit $z_\perp^i \rightarrow 0$ the expression simplifies further as, 
\begin{eqnarray}
&& \lim_{z_\perp^i \rightarrow 0} ~ \frac{\partial}{\partial z_\perp^i} ~ {\cal U}_{~[\eta]}\left(0,~z\right)  \nn \\
&& ~ \nn \\
&=& 
ig ~ U\left(0^-,~\eta^-;~0_\perp\right)
A^{i}\left(\eta^-,~0_\perp \right)
U\left(\eta^-,z^-;~0_\perp \right)  
 -~ig ~ U\left(0^-,~z^-;~0_\perp \right)  
A^{i}\left(z^-,~0_\perp  \right)    \nn \\
&&  +~ ig ~   \int^{\eta^-}_{ z^-}d\omega^- ~ U\left( 0^-,~\omega^-;~0_\perp  \right)
~ F^{+i}\left(\omega^-,~ 0_\perp \right)
~U\left(\omega^-, ~ z^-;~0_\perp \right)   ~. \label{LaLaLand3} \\
&& ~ \nn 
\end{eqnarray}
{\it Special Case}: ($\eta \rightarrow \pm \infty$)
\begin{eqnarray}
&& \left. \lim_{z_\perp^i \rightarrow 0} ~ \frac{\partial}{\partial z_\perp^i} ~ {\cal U}_{~[\eta = \pm \infty]}\left(0,~z\right) \right|_{\eta^- \rightarrow \pm\infty}  \nn \\
&& ~ \nn \\
&& ~ \nn \\
&&= ig ~ U\left(0^-,~\pm \infty^-;~0_\perp\right)
A^{i}\left(\pm \infty^-,~0_\perp \right)
U\left(\pm \infty^-,~z^-;~0_\perp \right)  
 -~ig ~ U\left(0^-,~z^-;~0_\perp \right)  
A^{i}\left(z^-,~0_\perp  \right)    \nn \\
&& ~ \nn \\
&&~~~~~~~~~~~~~~~~~~~~~~~  +~ ig ~   \int^{\pm \infty^-}_{ z^-}d\omega^- ~ U\left( 0^-,~\omega^-;~0_\perp  \right)
~ F^{+i}\left(\omega^-,~ 0_\perp \right)
~U\left(\omega^-, ~ z^-;~0_\perp \right)   ~. 
\label{}
\end{eqnarray} 
We can now express Eq.$\eqref{LaLaLand3}$ in terms of the physical gauge field $A^{i}_{\rm phys, \pm}$ and the pure gauge field $A^{i}_{\rm pure}$, 
\begin{eqnarray}
&& \lim_{z_\perp^i \rightarrow 0} ~ \frac{\partial}{\partial z_\perp^i} ~ {\cal U}_{[\eta]}\left(0,~z\right)  \nn \\
&& ~ \nn \\
&=& ig ~ U\left(0^-,~\eta^-;~0_\perp\right)
A^{i}\left(\eta^-,~0_\perp \right)
U\left(\eta^-,~z^-;~0_\perp \right)  
 -~ig ~ U\left(0^-,~z^-;~0_\perp \right)  
A^{i}\left(z^-,~0_\perp  \right)    \nn \\
&& ~ \nn \\
&&  -~ U\left(0^-,~\eta^-;~0_\perp\right)
A^{i}_{\rm phys, \pm}\left(\eta^-,~0_\perp \right)
U\left(\eta^-,~z^-;~0_\perp \right)
+  ig~U\left(0^-,~z^-;~0_\perp \right)  ~ A^{i}_{\rm phys,\pm}\left(z^-;~0_\perp \right)  \nn \\
&& ~ \nn \\
&=& 
ig ~ U\left(0^-,~\eta^-;~0_\perp\right)
A^{i}_{\rm pure}\left(\eta^-,~0_\perp \right)
U\left(\eta^-,~z^-;~0_\perp \right)  
 -~ig ~ U\left(0^-,~z^-;~0_\perp \right)  
A^{i}_{\rm pure}\left(z^-,~0_\perp  \right),      
\label{Joker2}
\end{eqnarray}
where we have used the definition of $A^{\mu}_{\rm phys}$ as in Eq.\eqref{LaLaLand4} and also Eq.\eqref{LaLaLand5}, 
which are based on the {\it Chen et. al.} decomposition \cite{Chen:2008ag, Chen:2009mr} of gauge fields as reviewed in the previous section. 
\section{Gluon Orbital Angular Momentum}

Orbital angular momentum of gluon can be expressed as the phase space average of the classical orbital angular momentum weighted with the Wigner distribution of polarized gluons in a longitudinally polarized nucleon.  
The Wigner distribution is essentially the phase space distribution of gluons in {\it transverse momentum} ($k_\perp$) - {\it impact parameter} ($b_\perp$) space, 
\begin{eqnarray}
{\cal W}_g \left(x, k_\perp, b_\perp \right)  
&=& \int \frac{d^2\Delta_\perp}{(2\pi)^2}~e^{-i\Delta_\perp b_\perp}
\int \frac{d^2z_\perp}{(2\pi)^2}~e^{iz_\perp k_\perp}
\int \frac{dz^-}{2(2\pi)~xP^+}~e^{-ixP^+ z^-}   \nn \\
&& ~~~~~ \times ~\left\langle P^+, ~-\frac{\Delta_\perp}{2},~ S \left| ~{\rm Tr} ~F^{+i}\left(0\right) ~ {\cal U}_{[\eta_1]}\left(0~;z\right)~ F^{+i}\left(z\right) ~ {\cal U}_{[\eta_2]}\left(z~;0\right)\right| P^+, ~\frac{\Delta_\perp}{2},~ S \right\rangle \label{Inception1} ~,
\end{eqnarray}
where $S$ is the spin of the target states, 
and the trace is in fundamental representation. 
The cross product of transverse momentum and impact parameter returns orbital angular momentum of gluons, integrating over them gives orbital angular momentum distribution, 
\begin{eqnarray}
L_g (x) &=& \int d^2b_\perp ~ d^2k_\perp  ~ \left(b_\perp \times k_\perp\right) ~ {\cal W}_g \left(x, k_\perp, b_\perp \right)   ~, \\
&=& \int d^2b_\perp ~ d^2k_\perp  ~ \epsilon^{kj}~b_\perp^k k_\perp^j ~ {\cal W}_g \left(x, k_\perp, b_\perp \right)  ~. 
\end{eqnarray}
This can further be written as, 
\begin{eqnarray}
L_g (x) &=& \int d^2b_\perp ~ \epsilon^{kj} ~ b_\perp^{k}~\int \frac{d^2\Delta_\perp}{(2\pi)^2}~e^{-i\Delta_\perp b_\perp}\int \frac{dz^-}{2(2\pi)~xP^+}~e^{-ixP^+ z^-}  ~  \nn \\
&& ~~~~~\times   \lim_{z_\perp^j \rightarrow 0}  \frac{1}{i} ~ \frac{\partial}{\partial z_\perp^j} ~~
\left\langle P^+, -\frac{\Delta}{2},~ S \left| ~{\rm Tr} ~~F^{+i}\left(0\right) ~ {\cal U}_{[\eta_1]}\left(0~;z\right)~ F^{+i}\left(z\right) ~ {\cal U}_{[\eta_2]}\left(z~;0\right) \right| P^+, +\frac{\Delta}{2}, ~S \right\rangle~. \label{Inception1}
\end{eqnarray}
Now the transverse derivative in Eq.\eqref{Inception1} can be performed using Eq.\eqref{Joker2} leading to, 
\begin{eqnarray}
L_g (x) &=& \int d^2b_\perp ~ \epsilon^{kj} ~ b_\perp^{k}~\int \frac{d^2\Delta_\perp}{(2\pi)^2}~e^{-i\Delta_i b_\perp}\int \frac{dz^-}{2(2\pi)~xP^+}~e^{-ixP^+ z^-} 
\frac{1}{i} ~
\left\langle P^+, -\frac{\Delta}{2}, S \left| ~{\rm Tr} ~ \sum_{m=1}^{5}~  {\cal O}_{m}^{j} ~ \right| P^+, +\frac{\Delta}{2}, S \right\rangle~.  \nn \\  \label{Inception12}
\end{eqnarray}

%
%
%
\noindent Here the object ${\cal O}_m$ are defined for convenience and are as follows, 
\begin{eqnarray}
{\cal O}_{1}^{j} &=&  F^{+i}\left(0\right)~ig~U\left(0^-,~\eta_1^-;~0_\perp\right)
A^{j}_{\rm pure} \left(\eta_1^-, ~ 0_\perp\right)  U\left(\eta_1^-,~z^-;~0_\perp\right) F^{+i}\left(z^-,~0_\perp\right){\cal U}_{[\eta_2]}\left(z^-,~0^-;~0_\perp\right), \nn \\
 \nn \\
{\cal O}_{2}^{j} &=&  -~F^{+i} \left(0\right)~ig~U\left(0^-,~z^-;~0_\perp\right)
A^{j}_{\rm pure}\left(z^-, ~ 0_\perp\right)  F^{+i}\left(z^-,~0_\perp\right)  {\cal U}_{[\eta_2]}\left(z^-,~0^-;~0_\perp\right), \nn \\
 \nn \\
{\cal O}_{3}^{j} &=&  F^{+i} \left(0\right)~{\cal U}_{[\eta_1]}\left(0^-,~z^-;~0_\perp\right) ~
\left[\partial^{j} F^{+i}\left(z^-,~z_\perp\right)\right]_{z_\perp =~ 0} ~ {\cal U}_{[\eta_2]}\left(z^-,~0^-;~0_\perp\right), \nn \\
 \nn \\
{\cal O}_{4}^{j} &=& F^{+i}\left(0\right) ~ {\cal U}_{[\eta_1]}\left(0~,z^-;~0_\perp\right)~ F^{+i}\left(z^-;~0_\perp\right) ~ ig ~ A^{j}_{\rm pure} \left(z^-,~0_\perp\right)~ U\left(z^-,~0^-;~0_\perp\right), \nn \\
 \nn \\
{\cal O}_{5}^{j} &=&  - F^{+i}\left(0\right) ~ {\cal U}_{[\eta_1]}\left(0~,z^-;~0_\perp\right)~F^{+i}\left(z^-;~0_\perp\right) ~ ig ~ U\left(z^-,~{\eta}^-_2;~0_\perp\right)~A^{j}_{\rm pure}\left({\eta_2}^-,~0_\perp\right)
~U\left({\eta}_2^-,~0^-;~0_\perp\right).     \nn
\end{eqnarray}

\noindent Eq. \eqref{Inception12} can further be simplified as, 
\begin{eqnarray}
L_g (x) &=& \epsilon^{kj}~\lim_{\Delta^k \rightarrow 0}\frac{\partial}{\partial \Delta^k}~ \int \frac{dz^-}{2(2\pi)xP^+}~e^{-ixP^+ z^-}  
~
\left\langle P^+, -\frac{\Delta}{2}, S\left| ~{\rm Tr} ~ \sum_{m=1}^{5}~  {\cal O}_{m}^{j} ~ \right| P^+, +\frac{\Delta}{2}, S \right\rangle  \nn \\  \label{Inception120}
\end{eqnarray}
Until now we have not taken any assumptions specifically for small-$x$. The problems now reduced to performing the $z^-$ integration. Now we expand the exponential in Eq.\eqref{Inception120}, 
\begin{eqnarray}
L_g (x) &=& \epsilon^{kj}~\lim_{\Delta^k \rightarrow 0}\frac{\partial}{\partial \Delta^k}~ \int \frac{dz^-}{2(2\pi)xP^+}~e^{-ixP^+ z^-}  
~
\left\langle P^+, -\frac{\Delta}{2}, S \left| ~{\rm Tr} ~ \sum_{m=1}^{5}~  {\cal O}_{m}^{j} ~ \right| P^+, +\frac{\Delta}{2}, S \right \rangle  \nn \\  
&=& \epsilon^{kj}~\lim_{\Delta^k \rightarrow 0}\frac{\partial}{\partial \Delta^k}~ \int \frac{dz^-}{2(2\pi)}~\sum_{n=0}^{\infty} (-i)^n (xP^+)^{n-1} {z^-}^n
~
\left\langle P^+, -\frac{\Delta}{2}, S \left| ~{\rm Tr} ~ \sum_{m=1}^{5}~  {\cal O}_{m}^{j} ~ \right| P^+, +\frac{\Delta}{2}, S \right\rangle  \nn \\
&\equiv&\epsilon^{kj}~ \left[ \frac{1}{x}~{\cal L}^{jk}_{g,\rm {0}}\left(x\right) + {\cal L}^{jk}_{g,\rm {1}}\left(x\right) +  x~{\cal L}^{jk}_{g,\rm {2}}\left(x\right) + x^2  {\cal L}^{jk}_{g,\rm {3}}\left(x\right) + ...   \right]
\label{Inception121}
\end{eqnarray}
Here ${\cal L}^{jk}_{g,\rm {0}}\left(x\right)$, ${\cal L}^{jk}_{g,\rm {1}}\left(x\right)$, ${\cal L}^{jk}_{g, \rm {2}}\left(x\right)$ ... can be interpreted as eikonal, sub-eikonal, sub-sub-eikonal contributions to the gluon OAM. 
In this convention, sub-eikonal refers to the term suppressed by one power of $x$ compared to the eikonal or leading scattering, sub-sub-eikonal refers to suppression by two power of $x$, and so on. 
Any dependence on $x$, for the terms, ${\cal L}^{jk}_{g,\rm {0}}\left(x\right)$, ${\cal L}^{jk}_{g,\rm {1}}\left(x\right)$, ${\cal L}^{jk}_{g, \rm {2}}\left(x\right)$ ..., can be estimated only by constructing and solving small-$x$ evolution of the  object inside the angle bracket $<...>$. In the following we will study the operator structure of eikonal and sub-eikonal terms in Eq.\eqref{Inception121}.

\section{Gluons OAM at eikonal limit}
In this leading order we can approximate 
$\exp\left(-ixP^+z^-\right) \simeq 1$, and go ahead to perform the $z^-$ integration in 
Eq.\eqref{Inception120}. 
%
Now its also important to isolate only the polarised contributions, from the object inside the angle bracket $<...>$, that would give one more $\epsilon^{ij}$, in addition to the one that is already there originating from the definition of OAM, leading to surviving OAM for gluons. 
%
%
%
%
%
%
Now before we proceed, it will be convenient to define the following PT even and PT odd parts of $A_{\rm phys}$, 
\begin{eqnarray}
A^{\mu}_{\rm phys,~ e}(x) &=& \frac{1}{2} \left[A^{\mu}_{\rm phys,+}(x)+A^{\mu}_{\rm phys,-}(x)\right]  \nn \\
&=& -\frac{1}{2}\int_{-\infty^-}^{+\infty^-}dy^- \epsilon(y^-)~ U(x^-,y^-;~x_\perp)~F^{+\mu}(y^-, ~x_\perp) ~U(y^-,x^-;~x_\perp) \label{PTeven} \\
&~&   \nn \\
&~&   \nn \\
A^{\mu}_{\rm phys,~ o}(x) &=& \frac{1}{2} \left[A^{\mu}_{\rm phys,+}(x)-A^{\mu}_{\rm phys,-}(x)\right]  \nn \\
&=& -\frac{1}{2}\int_{-\infty^-}^{+\infty^-}dy^- ~ U(x^-,y^-;~x_\perp)~F^{+\mu}(y^-, ~x_\perp)~ U(y^-,x^-;~x_\perp) \label{PTodd}
\end{eqnarray}
with $\epsilon(y^-)$ being the sign function. Now here we list the integral over $z^-$ of the respective ${\cal O}_m$, details of $z^-$ integrations are in the appendix,  

\begin{eqnarray}
\int dz^-~{\cal O}_{1}^{j} &=& -2ig~F^{+i}\left(0^-,~0_\perp\right) ~ U\left(0^-,~\eta_1^-;~0_\perp\right) ~ A^j_{\rm pure}\left(\eta_1^-,~0_\perp\right) ~ U\left(\eta_1^-,~\eta_2^-;~0_\perp\right) ~ A_{\rm phys,~ o}^i\left(\eta^-_2,~0_\perp\right)~U\left(\eta_2^-,~0^-; ~ 0_\perp\right) \nn \\
\nn \\
\int dz^-~{\cal O}_{2}^{j}  &=&  2ig~F^{+i}\left(0^-,~0_\perp\right)
A^{j}_{\rm res}\left(0^-,~0_\perp\right)   ~ A^{i}_{\rm phys,~ o}\left(0^-,~0_\perp\right)    \nn \\
\nn \\
\int dz^-~{\cal O}_{3}^{j}  &=&  -2F^{+i}\left(0^-,~0_\perp\right) ~ \partial^j 
A^{i}_{\rm phys,~ o}\left(0^-,~0_\perp\right)  \nn \\
\nn \\
\int dz^-~{\cal O}_{4}^{j}  &=& -2ig~F^{+i}\left(0^-,~0_\perp\right) ~ A_{\rm phys,~ o}^{i}\left(0^-,~0_\perp\right)~ A_{\rm res}^{j}\left(0^-,~0_\perp\right)  \nn \\
\nn \\
\int dz^-~{\cal O}_{5}^{j}  &=& 2ig~F^{+i}\left(0^-,~0_\perp \right)  U\left(0^-,~\eta_1^-;~0_\perp\right)  A^{i}_{\rm phys,~ o}\left(\eta_1^-,~0_\perp\right) U\left(\eta_1^-,
~{\eta_2^-};~0_\perp\right) A_{\rm pure}^j ~\left(\eta_2^-,~0_\perp\right) ~U\left(\eta_2^-,~0^-;~0_\perp\right)   \nn 
\end{eqnarray}
\vspace{0.7cm}

\noindent This results can be supplemented back to Eq.\eqref{Inception120} to arrive at the final expression of the Gluon OAM at small-$x$,
\begin{eqnarray}
{\cal L}^{jk}_{g,\rm {0}}\left(x\right) &=& ~-\lim_{\Delta^k \rightarrow 0}\frac{\partial}{\partial \Delta^k}~ \left\langle P^+, -\frac{\Delta}{2},~ S \left| ~{\rm Tr} ~ F^{+i}\left(0^-,~0_\perp\right) ~ \partial^j A^{i}_{\rm  phys,~ o}\left(0^-,~0_\perp\right) ~ \right| P^+, +\frac{\Delta}{2}, ~S \right\rangle  \nn \\
\nn \\
&& ~- \lim_{\Delta^k \rightarrow 0}\frac{\partial}{\partial \Delta^k}~ \left\langle P^+, -\frac{\Delta}{2},~ S \left| ~{\rm Tr} ~ F^{+i}\left(0^-,~0_\perp\right) ~ig~ \left[A^i_{\rm  phys,~ o}~, ~A^j_{\rm res}\right] \left(0^-,~0_\perp\right) ~ \right| P^+, +\frac{\Delta}{2}, ~S \right\rangle  \nn \\
\nn \\
&& ~- \lim_{\Delta^k \rightarrow 0}\frac{\partial}{\partial \Delta^k}~ \left\langle P^+, -\frac{\Delta}{2},~ S \left| ~   ig F^{+i}(0^-,0_\perp)~U\left(0^-, \eta^-_1; 0_\perp \right)\left[ A^j_{\rm pure}\left(\eta_1^-, 0_\perp \right) U(\eta_1^-, \eta_2^-; 0_\perp)A^i_{\rm phys}\left(\eta_2^-,0_\perp\right)  \right.  \right. \right. \nn\\
&&\left. ~~~~~~~~~~~~~~~~~~~~~~~~~~~~~  \left. \left. -A^{i}_{\rm phys}\left(\eta_1^-,0_\perp \right) U(\eta_1^-, \eta_2^-;0_\perp) A^j_{\rm pure}\left(\eta_2^-, 0_\perp\right)\right] U(\eta_2^-,0^-;0_\perp) \right| P^+, +\frac{\Delta}{2}, ~S \right\rangle    
\label{Godfather4}
\end{eqnarray}
\vspace{0.7cm}

\noindent In the contour gauge e.g. in $A^+=0$ gauge, 
$A^{\rm res} = A^{\rm pure}$, which leads to, 
\begin{eqnarray}
\left[A^{i}_{\rm phys,~o},~ A^{j}_{\rm res}\right] \mapsto \left[A^{i}_{\rm phys,~o},~ A^{j}_{\rm pure}\right]
\end{eqnarray}
\vspace{0.7cm}

\noindent The eikonal contribution to gluon OAM operator, ${\cal L}^{g}_{\rm {0}}(x)$, using Eq.\eqref{PTodd}, can then be written as, 
 \begin{eqnarray}
&&{\cal L}^{jk}_{g,\rm {0}}(x, k_\perp, \Delta_\perp, S) \nn \\
&=& -~\frac{1}{2}\lim_{\Delta^k \rightarrow 0}\frac{\partial}{\partial \Delta^k}~ \left\langle P^+, -\frac{\Delta}{2},~ S \left| ~{\rm Tr} ~ F^{+i}\left(0^-,~0_\perp\right) ~ {\cal D}^j_{\rm pure} \left(A^{i}_{\rm  phys,+}-A^{i}_{\rm  phys,-}\right)\left(0^-,~0_\perp\right) ~ \right| P^+, +\frac{\Delta}{2}, ~S \right\rangle  \nn \\
\nn \\
&& - ~ \frac{1}{2} \lim_{\Delta^k \rightarrow 0}\frac{\partial}{\partial \Delta^k}~ \left\langle P^+, -\frac{\Delta}{2},~ S \left| ~   ig F^{+i}(0^-,0_\perp)~U\left(0^-, \eta^-_1; 0_\perp \right) 
\left[ A^j_{\rm pure}\left(\eta_1^-, 0_\perp \right) U(\eta_1^-, \eta_2^-; 0_\perp)  \right.  \right. \right. \nn\\
\nn \\
&&\left. ~ ~ ~ \left(A^{i}_{\rm  phys,+}-A^{i}_{\rm  phys,-}\right)\left(\eta_2^-,0_\perp\right) \left. \left. -\left(A^{i}_{\rm  phys,+}-A^{j}_{\rm  phys,-}\right)\left(\eta_1^-,0_\perp \right) U(\eta_1^-, \eta_2^-;0_\perp)A^j_{\rm pure} \left(\eta_2^-, 0_\perp\right)\right] \right. \right. \nn \\
\nn \\
&&\left. \left. ~~~ U(\eta_2^-,0^-;0_\perp) \right| P^+, +\frac{\Delta}{2}, ~S \right\rangle,   \label{Godfather}
\end{eqnarray}
where we have explicitly shown the presence of $A^{i}_{\rm phys,\pm}$ in the expression. Now under PT transformation, 
\begin{eqnarray}
\Delta_\perp & \rightarrow & -\Delta_\perp, \nn \\
S & \rightarrow & - S, \nn \\ 
F^{\mu\nu} & \rightarrow & - F^{\mu\nu}, \nn \\
A^{\mu}_{\rm phys,\pm} & \rightarrow & A^{\mu}_{\rm phys,\mp}~.
\end{eqnarray}
This implies that ${\cal L}^{jk}_{g,\rm {0}}(x, k_\perp, \Delta_\perp, S)$ and its PT transformed pair are identical $i.e.$, 
\begin{eqnarray}
{\cal L}^{jk}_{g,\rm {0}}(x, k_\perp, \Delta_\perp, S) = {\cal L}^{jk}_{g,\rm {0}}(x, -k_\perp, -\Delta_\perp, -S),
\end{eqnarray}
which ensure that the off-forward matrix elements, in ${\cal L}^{jk}_{g,\rm {0}}(x, k_\perp, \Delta_\perp, S)$, do not have any spin dependent structure of the following form,
\begin{eqnarray}
i\frac{S^+}{P^+} \epsilon^{lm} \Delta^l_\perp k^m_\perp ...
\end{eqnarray}
In absence of another totally antisymmetric tensor $\epsilon$ inherent to ${\cal L}^{jk}_{g,\rm {0}}$, the eikonal contribution in Eq.\eqref{Inception121} vanishes as, 
\begin{eqnarray}
\epsilon^{kj}{\cal L}^{jk}_{g,\rm {0}} = 0.
\end{eqnarray}
Non-vanishing spin effect starts showing up only at sub eikonal level which we will discuss next.

\section{Gluon OAM at first sub-eikonal order}
\noindent When one moves to next non-trivial order in the expansion of the exponential,  
as $\exp\left(-ixP^+z^-\right) \simeq 1- ixP^+ z^-$, the additional $i$ makes all leading T-even terms to be T-odd, leading to non-zero contribution to gluon OAM for longitudinally polarized targets, 
\begin{eqnarray}
{\cal L}^{jk}_{g,\rm {1}}\left(x\right) &=& ~iP^+\lim_{\Delta^k \rightarrow 0}\frac{\partial}{\partial \Delta^k}~ \left\langle P^+, -\frac{\Delta}{2},~ S \left| ~{\rm Tr} ~ F^{+i}\left(0^-,~0_\perp\right) ~ \partial^j {\bar O}^{i}_{\rm phys,*}\left(0^-,~0_\perp\right) ~ \right| P^+, +\frac{\Delta}{2}, ~S \right\rangle  \nn \\
\nn \\
&& + iP^+\lim_{\Delta^k \rightarrow 0}\frac{\partial}{\partial \Delta^k}~ \left\langle P^+, -\frac{\Delta}{2},~ S \left| ~{\rm Tr} ~ F^{+i}\left(0^-,~0_\perp\right) ~ig~ \left[{\bar O}^{i}_{\rm phys,*}, ~A^j_{\rm res}\right] \left(0^-,~0_\perp\right) ~ \right| P^+, +\frac{\Delta}{2}, ~S \right\rangle  \nn \\
\nn \\
&& +iP^+\lim_{\Delta^k \rightarrow 0}\frac{\partial}{\partial \Delta^k}~ \left\langle P^+, -\frac{\Delta}{2},~ S \left| ~   ig F^{+i}(0^-,0_\perp)~U\left(0^-, \eta^-_1; 0_\perp \right)\left[ A^j_{\rm pure}\left(\eta_1^-, 0_\perp \right) U(\eta_1^-, \eta_2^-; 0_\perp)~{\bar O}^{i}_{\rm phys,*}\left(\eta_2^-,0_\perp\right)  \right.  \right. \right. \nn\\
&&\left. ~~~~~~~~~~~~~~~~~~~~~~  \left. \left. -{\bar O}^{i}_{\rm phys,*}\left(\eta_1^-,0_\perp \right) U(\eta_1^-, \eta_2^-;0_\perp) A^j_{\rm pure}\left(\eta_2^-, 0_\perp\right)\right] U(\eta_2^-,0^-;0_\perp) \right| P^+, +\frac{\Delta}{2}, ~S \right\rangle    
\label{Godfather5}
\end{eqnarray}
where, 
 \begin{eqnarray}
{\bar O}^{i}_{\rm phys,*}(x) 
&=& -\frac{1}{2}\int_{-\infty^-}^{+\infty^-}dz^- ~ z^-~ U(x^-,z^-;~x_\perp)~F^{+i}(z^-, ~x_\perp)~ U(z^-,x^-;~x_\perp). \label{AA01}
\end{eqnarray}
This is essentially the $z^-$ moment of the physical gauge $A^{i}_{\rm phys, o}$ in Eq.\eqref{Godfather}. 
However, unlike $A^{i}_{\rm phys, o}$, the object ${\bar O}^{i}_{\rm phys,*}$ in the above expression is PT even, similar to Eq.\eqref{PTeven}, because of the presence of an extra $z^-$, in the integrand, which can be written as $z^-=|z^-|~{\rm sign}(z^-)$.  This implies that unlike $ {\cal L}^{jk}_{g,\rm {0}}$, the term ${\cal L}^{jk}_{g,\rm {1}}(x, k_\perp, \Delta_\perp, S)$ is odd under PT transformation as,  
\begin{eqnarray}
{\cal L}^{jk}_{g,\rm {1}}(x, k_\perp, \Delta_\perp, S) = - {\cal L}^{jk}_{g,\rm {1}}(x, -k_\perp, -\Delta_\perp, -S).
\end{eqnarray}
One would expect to get non-trivial contribution to gluon OAM, in longitudinally polarized proton, right from this order.
As the expression is originally stems from a Fourier transform, one can get the intermediate time $z^-$ out of the integral to have an overall energy derivative, 
 \begin{eqnarray}
{\bar O}^{i}_{\rm phys,*}(x) 
&=& -\frac{1}{2} ~i\partial_{k}^{+}\int_{-\infty^-}^{+\infty^-}dz^- ~ U(x^-,z^-;~x_\perp)~F^{+i}(z^-, ~x_\perp)~ U(z^-,x^-;~x_\perp),  \nn \\
&=&  -\frac{1}{2}~ i\partial_{k}^{+} \left(A^{i}_{\rm  phys,+}-A^{i}_{\rm  phys,-}\right),
\label{AA02}
\end{eqnarray}
which leads to, 
 \begin{eqnarray}
&&{\cal L}^{jk}_{g,\rm {1}}(x, k_\perp, \Delta_\perp, S) \nn \\
&=& \frac{1}{2}\lim_{\Delta^k \rightarrow 0}\frac{\partial}{\partial \Delta^k}\frac{\partial}{\partial x}~ \left\langle P^+, -\frac{\Delta}{2},~ S_\perp \left| ~{\rm Tr} ~ F^{+i}\left(0^-,~0_\perp\right) ~ {\cal D}^j_{\rm pure} \left(A^{i}_{\rm  phys,+}-A^{i}_{\rm  phys,-}\right)\left(0^-,~0_\perp\right) ~ \right| P^+, +\frac{\Delta}{2}, ~S_\perp \right\rangle  \nn \\
\nn \\
&& +  \frac{1}{2} \lim_{\Delta^k \rightarrow 0}\frac{\partial}{\partial \Delta^k}\frac{\partial}{\partial x}~ \left\langle P^+, -\frac{\Delta}{2},~ S_\perp \left| ~   ig F^{+i}(0^-,0_\perp)~U\left(0^-, \eta^-_1; 0_\perp \right) 
\left[ A^j_{\rm pure}\left(\eta_1^-, 0_\perp \right) U(\eta_1^-, \eta_2^-; 0_\perp)  \right.  \right. \right. \nn\\
\nn \\
&&\left. ~ ~ ~ \left(A^{i}_{\rm  phys,+}-A^{i}_{\rm  phys,-}\right)\left(\eta_2^-,0_\perp\right) \left. \left. -\left(A^{i}_{\rm  phys,+}-A^{j}_{\rm  phys,-}\right)\left(\eta_1^-,0_\perp \right) U(\eta_1^-, \eta_2^-;0_\perp)A^j_{\rm pure} \left(\eta_2^-, 0_\perp\right)\right] \right. \right. \nn \\
\nn \\
&&\left. \left. ~~~ U(\eta_2^-,0^-;0_\perp) \right| P^+, +\frac{\Delta}{2}, ~S_\perp \right\rangle.   \label{Godfather}
\end{eqnarray}
This also corroborate the fact that in the Taylor expansion of the phase factor $\exp(ixP^+z^-)$, only the odd terms in $x$ can contribute to the gluon OAM for longitudinally polarized proton \cite{Hatta:2016aoc}. 

\section{conclusion}
Recently the connection between gluon Wigner distribution and gluon orbital angular momentum (OAM) have been used to probe the gluon OAM in the hard scattering process at the Electron~-~Ion Collider. The single longitudinal spin asymmetry in hard diffractive dijet production  found to be sensitive to the gluon orbital angular momentum distribution, at least in the moderate $x$-range \cite{Ji:2016jgn}. 
In the theoretical front, specifically at small-$x$, 
Hatta presented a general analysis of the orbital angular momentum  of distribution of gluons \cite{Hatta:2011ku,Hatta:2016aoc}. Novel operator representation of $L_g(x)$ has been derived at small-$x$ limit and interestingly found to contain covariant derivatives inserted at some intermediate time between far past and far future. 
%
Moreover, the orbital angular momentum distribution was found to be proportional to the gluon helicity distributions $L_g(x)\approx -\Delta G(x)$ at small-$x$.  
 Subsequently, Kovchegov derived the small-$x$ asymptotics of the gluon orbital angular momentum distribution  of proton in the double logarithmic approximation \cite{Kovchegov:2019rrz}. The procedure adopted was to start with the operator definition of gluon OAM, simplifying it at small-$x$ and relating it to the polarized dipole amplitudes for the gluon helicities. The small-$x$ asymptotic of the later then utilized to derive the small-$x$ asymptotic of the OAM at large $N_c$ limit. 
\vspace{0.4cm}


 In this work we  studied the transverse derivatives of the staple gauge links, with varying extent along the light cone. 
 This is partly motivated by the recent work by M. Engelhardt \cite{Engelhardt:2020qtg}
where the authors considered a general form of staple shaped path and computed both Ji and Jaffe-Manohar orbital angular momentum in lattice using the direct derivative method. The derivative of such gauge link with respect to transverse position at zero transverse position limit has been calculated and expressed in terms of the pure gauge components of the gauge fields within the framework of Chen et. al. decomposition of gauge fields. 
\vspace{0.4cm}

Wigner distribution of the gluon is essentially the  distribution of gluons in impact parameter and transverse momentum space. The connection between the two has been recalled to derive the OAM distribution in terms of the pure or physical gauge components and the gauge link parameters. After performing the integration over both the $z^-$ and $z_\perp$ we derive the general operator form of the OAM, of gluons in a longitudinally polarized proton, that is valid for all possible geometrics of the gauge links. At an appropriate combination of the extent parameters this correctly reproduces both Jaffe-Manohar and Ji's OAM, and offers a continuous analytical interpolation between the two, for gluon OAM distribution in a longitudinally polarised proton. 


 \begin{acknowledgments}
 \noindent We sincerely acknowledge and thank the anonymous referee for the critical reading and all the suggestions that substantially improve the article. 
The work is supported by the DST - Govt. of India through SERB-MATRICS Project Grant No. MTR/2019/001551. 
 \end{acknowledgments}

\section{appendix}

\subsection{Integration over $z^-$}

\subsubsection{For ${\cal O}_1^{j}$ and ${\cal O}_5^{j}$}

\begin{eqnarray}
\int dz^-~{\cal O}_{1}^{j}   &=&  \int dz^- F^{+i}\left(0\right)~ig~U\left(0^-,~\eta_1^-;~0_\perp\right)
A^{j}_{\rm pure} \left(\eta_1^-, ~ 0_\perp\right)  U\left(\eta_1^-,~z^-;~0_\perp\right) F^{+i}\left(z^-,~0_\perp\right){\cal U}_{[\eta_2]}\left(z^-,~0^-;~0_\perp\right)   \nn
\end{eqnarray}

\vspace{0.7cm}
\noindent Now, 
\begin{eqnarray}
{\cal U}_{[\eta_2]}\left(z^-,~0^-;~0_\perp\right) = {U}_{}\left(z^-,~0^-;~0_\perp\right)
\end{eqnarray}
and, 
\begin{eqnarray}
&&~~~~~ \int_{-\infty}^{+\infty} dz^- ~U\left(\eta_2^-,~z^-;~0_\perp\right) F^{+i}\left(z^-,~0_\perp\right){ U}_{}\left(z^-,~\eta_2^-;~0_\perp\right) 
\nn \\
\nn \\
&& = -\int_{+\infty}^{\eta_2^-} dz^- ~U\left(\eta_2^-,~z^-;~0_\perp\right) F^{+i}\left(z^-,~0_\perp\right){U}_{}\left(z^-,~\eta_2^-;~0_\perp\right) \nn \\
&& ~~~~~~~~~~+
\int_{-\infty}^{\eta_2^-} dz^- ~U\left(\eta_2^-,~z^-;~0_\perp\right) F^{+i}\left(z^-,~0_\perp\right){U}_{}\left(z^-,~\eta_2^-;~0_\perp\right) 
\nn \\
\nn \\
&& = -A_{\rm phys, +}\left(\eta_2^-,~0_\perp\right) + A_{\rm phys, -}\left(\eta_2^-,~ 0_\perp\right)
\nn \\
\nn \\
&&=  -2A_{\rm phys,~o}\left(\eta_2^-,~0_\perp\right) \nn
\end{eqnarray}
Therefore, 
\begin{eqnarray}
\int dz^-~{\cal O}_{1}^{j}   &=& -2ig~F^{+i}\left(0^-,~0_\perp\right) ~ U\left(0^-,~\eta_1^-;~0_\perp\right) ~ A^j_{\rm pure}\left(\eta_1^-,~0_\perp\right) ~ U\left(\eta_1^-,~\eta_2^-;~0_\perp\right) ~ A_{\rm phys,~ o}^i\left(\eta^-_2,~0_\perp\right)~U\left(\eta_2^-,~0^-; ~ 0_\perp\right) \nn 
\end{eqnarray}
Similarly one can get, 
\begin{eqnarray}
\int dz^-~{\cal O}_{5}^{j}   &=& - \int dz^- F^{+i}\left(0\right) ~ {\cal U}_{[\eta_1]}\left(0~,z^-;~0_\perp\right)~F^{+i}\left(z^-;~0_\perp\right) ~ ig ~ U\left(z^-,~{\eta}^-_2;~0_\perp\right)~A^{j}_{\rm pure}\left({\eta_2}^-,~0_\perp\right)
~U\left({\eta}_2^-,~0^-;~0_\perp\right)   \nn \\
 \nn \\
&=& 2ig~F^{+i}\left(0^-,~0_\perp \right)  U\left(0^-,~\eta_1^-;~0_\perp\right)  A^{i}_{\rm phys,~ o}\left(\eta_1^-,~0_\perp\right) U\left(\eta_1^-,
~{\eta_2^-};~0_\perp\right) A_{\rm pure}^j ~\left(\eta_2^-,~0_\perp\right) ~U\left(\eta_2^-,~0^-;~0_\perp\right)  \nn 
\end{eqnarray}

\vspace{1cm}
\subsubsection{For ${\cal O}_2^{j}$ and ${\cal O}_4^{j}$}

\begin{eqnarray}
\int dz^-~{\cal O}_{2}^{j}   &=& - \int dz^- ~ F^{+i} \left(0^-,0_\perp\right)~ig~U\left(0^-,~z^-;~0_\perp\right)
A^{j}_{\rm pure}\left(z^-, ~ 0_\perp\right)  F^{+i}\left(z^-,~0_\perp\right)  {\cal U}_{[\eta_2]}\left(z^-,~0^-;~0_\perp\right) \nn 
\end{eqnarray}
In the first step we perform the following transformation of 
$A_{\rm pure}$ at $(z^-,~0_\perp)$ to the reference point $(0^-,~0_\perp)$. 
\begin{eqnarray}
U(0^-,z^-;0_\perp) ~A^j_{\rm pure}\left(z^-, 0_\perp\right)~U\left(z^-,0^-; ~0_\perp\right) \mapsto A^j_{\rm res}\left(0^-, 0_\perp\right) \nn 
\end{eqnarray}
and then perform the $z^-$ integration, 
\begin{eqnarray}
&& \int_{-\infty}^{+\infty} dz^- ~U\left(0^-,~z^-;~0_\perp\right) F^{+i}\left(z^-,~0_\perp\right)U\left(z^-,~0^-;~0_\perp\right)   \nn \\
\nn \\
&& = -2A_{\rm phys,~o}\left(0^-, ~0_\perp\right)  \nn 
\end{eqnarray}
which leads to, 
\begin{eqnarray}
\int dz^-~{\cal O}_{2}^{j}  =   2ig ~F^{+i} \left(0^-,~0_\perp\right) A^j_{\rm res}\left(0^-, 0_\perp\right)   A_{\rm phys,~o}\left(0^-, ~0_\perp\right), \nn 
\end{eqnarray}
Similarly, 
\begin{eqnarray}
\int dz^-~{\cal O}_{4}^{j}  =   -2ig ~F^{+i} \left(0^-,~0_\perp\right)A_{\rm phys,~o}\left(0^-, ~0_\perp\right)
 A^j_{\rm res}\left(0^-, 0_\perp\right).  \nn
\end{eqnarray}

\subsubsection{For ${\cal O}_3^{j}$}
\begin{eqnarray}
\int dz^-~{\cal O}_3^{j} &=&  \int_{-\infty}^{+\infty} dz^- ~ F^{+i} \left(0\right)~
{\cal U}_{[\eta_1]}\left(0^-,~z^-;~0_\perp\right) ~
\left[\partial^{j} F^{+i}\left(z^-,~z_\perp\right)\right]_{z_\perp =~ 0} ~ {\cal U}_{[\eta_2]}\left(z^-,~0^-;~0_\perp\right) \nn 
\end{eqnarray}
In order to perform integration we shall first change the order of the integration and differentiation. This is legit as both are linear operators. 
This will lead to, 
\begin{eqnarray}
\int dz^-~{\cal O}_3^{j} &=&    -2F^{+i} \left(0^-, ~0_\perp\right) ~
\partial^{j} A^{i}_{\rm phys, ~o}\left(0^-,~0_\perp\right).  \nn 
\end{eqnarray}

 \end{document}